\documentclass[aps,pra,preprint,superscriptaddress,showpacs,showkeys]{revtex4}
\begin{document}
\title{Paraxial spin transport using the Dirac-like paraxial wave equation}

\author{Mohammad Mehrafarin}
\email{mehrafar@aut.ac.ir}
\affiliation{Physics Department, Amirkabir University of Technology, Tehran 15914, Iran}
\author{Hamideh Balajany}
\affiliation{Physics Department, Alzahra University, Tehran 19938, Iran}

\begin{abstract}
In weakly inhomogeneous media, Maxwell equations assume a Dirac-like form that is particularly apt for the study of paraxial propagation. Using this form, and via the Foldy-Wouthuysen transformation technique of the Dirac equation, we study the spin transport of paraxial light beams in weakly inhomogeneous media. We derive the Berry effect terms and establish the spin Hall effect and the Rytov rotation law for polarized paraxial beam transport.
\end{abstract}

\maketitle

\section{Introduction}

The first consistent analysis of the geometrical optics approximation for electromagnetic waves in weakly inhomogeneous isotropic media was reported in a work by Rytov \cite{Rytov}. He studied the geometric optics to first approximation in the wave number and gave his
famous law of the rotation of the electromagnetic wave polarization plane. Geometrical attributes of this law were described further in detail by Vladimirskii \cite{Vladimirskii}. The Rytov-Vladimirskii rotation of the polarization plane, which was observed experimentally in the 1980's \cite{Tomito}, has been shown (see \cite{Shapere} and the references therein) to be an example of the geometric Berry phase \cite{Berry}. Another fundamental phenomenon associated with the propagation of light in weakly inhomogeneous media was discovered by Zel'dovich and co-workers in the 1990's \cite{Zeldovich,Zeldovich2}. It was termed the `optical Magnus effect', according to which, waves of right and left circular polarization propagate along different trajectories. 

A decade later, Bliokh and co-workers \cite{Bliokh,Bliokh1,Bliokh2} developed a modified geometrical optics approximation for weakly inhomogeneous isotropic media, which contained the Rytov law and the optical Magnus effect as manifestations of Berry effects. In this approach, a Berry connection (gauge potential) in momentum space stems from the representation in which the wave equation is diagonal. The Berry curvature (gauge field strength) associated with this connection corresponds to a magnetic-monopole-like field. The connection results in a Berry phase that describes the Rytov rotation of the polarization plane. The curvature causes the optical Magnus effect, the deviations of waves of right/left polarization being in opposite directions and orthogonal to the direction of propagation. This produces a spin (polarization) current directed across the principal motion of the beam. Such geometric Berry effects have been derived via other approaches too \cite{Onoda,Duval}. The situation is similar to the anomalous and the spin Hall effect of electrons \cite{Culcer,Murakami,Sinova,Kato,Horvathy} and is, therefore, also referred to as the optical Hall effect or the spin Hall effect of photons.

Because of their capacity to carry orbital angular momentum in addition to spin, paraxial light beams have found considerable interest  \cite{Allen,Beijersbergen,Simpson,Allen2,Maier,Piccirillo}. The spin transport of such beams has been studied via the standard paraxial wave equation with emphasis on the orbital-angular-momentum Hall effect \cite{Bliokh3,Bliokh4}. Now, in weakly inhomogeneous media, where permittivity varies adiabatically with position (the permeability being constant), Maxwell equations assume a Dirac-like form that is particularly apt for the study of paraxial propagation (see next section). In the present work, we take advantage of this Dirac-like form to study paraxial spin transport in weakly inhomogeneous media. Our approach is similar to that used for relativistic electrons \cite{Berard} via the Foldy-Wouthuysen transformation technique which diagonalizes the Dirac Hamiltonian. We, thus, derive the Berry effect terms and establish the spin Hall effect and the Rytov rotation law for polarized paraxial beam transport. 

\section{Maxwell equations in Dirac-like form}

In a free weakly inhomogeneous medium, where the (relative) permittivity $\epsilon(\textbf{x})$ varies adiabatically with position ($\nabla\epsilon \rightarrow 0$), Maxwell equations can be written as \cite{Khan}
\begin{equation}
\frac{n}{c}\:\partial_t\Psi=-\textbf{M}\cdot\nabla \Psi \label{1}
\end{equation}
$n(\textbf{x})=\surd {\epsilon(\textbf{x})}$ being the refractive index. Here,
\begin{equation}
\Psi(\textbf{x},t)=\left( \begin{array}{c}-F_{x}+iF_{y}\\ F_{z}\\ F_{z}\\F_{x}+ iF_{y} \end{array} \right) \label{0} 
\end{equation}
where $\textbf{F} (\textbf{x},t)=n(\textbf{x})\textbf{E}(\textbf{x},t) +ic \textbf{B}(\textbf{x},t)$ is the Reimann-Silberstein vector and
\begin{equation}
M_{x}=\left( \begin{array}{cc}0&I\\I&0\end{array} \right), \ \
M_{y}=i\left( \begin{array}{cc}0&-I\\I&0\end{array} \right),\ \
M_{z}=\left( \begin{array}{cc}I&0\\0&-I\end{array} \right)=\beta \label{3}
\end{equation}
$I$ being the $2\times2$ unit matrix. The matrices $M_i$ are Hermitian and satisfy the algebra of the Dirac matrices, including
\begin{equation}
M_i^2=1, \ \ M_iM_j=-M_jM_i=iM_k \ (i,j,k\ \text{cyclic}). \label{a}
\end{equation}
For monochromatic waves $\Psi(\textbf{x},t)=e^{-i\omega t}\Psi_0(\textbf{x})$, Maxwell equations (\ref{1}) reduce to
\begin{equation}
(\textbf{M}\cdot\textbf{p})\Psi_0=n \Psi_0 \label{2}
\end{equation}
where $\textbf{p}=-ik^{-1}\nabla$ is the (dimensionless) `momentum' operator, $k=\omega/c$ being the wave number. $\textbf{p}$ represents the momentum (unit wave vector) of the wave and satisfies the standard commutation relations $[x_i,p_j]=ik^{-1}\delta_{ij}$. (Without loss of clarity, we shall use the same symbol for an operator and its eigenvalue to simplify the notation.) In the adiabatic approximation, let us write $n(\textbf{x})=n_0+\zeta(\textbf{x})$, where $\zeta(\textbf{x})$ is a small perturbation around the uniform background value $n_0$. Introducing the notation $\textbf{M}_\bot=(M_x,M_y,0)$, $\textbf{p}_\bot=(p_x,p_y,0)$, etc., equation (\ref{2}) may be written as the Dirac-like equation
\begin{equation}
ik^{-1}\partial_z \Psi_0=H \Psi_0 \label{dirac} 
\end{equation}
where 
$$H=-n_0 \beta-\zeta\beta+\beta \textbf{M}_\bot \cdot \textbf{p}_\bot=-n_0 \beta-\zeta\beta+i\textbf{M}_\bot \cdot (\hat{\textbf{z}} \times \textbf{p}_\bot)$$ 
having used (\ref{a}) in the last expression. Note, however, that the `Hamiltonian' $H$ is not Hermitian because of the last term, which we shall denote by $O$. The above form is particularly apt for the paraxial approximation, with $z$ as the paraxial direction and the subscript $\bot$ signifying the transverse plane. In the paraxial approximation we have $k_\bot=\sqrt{k_x^2+k_y^2}\ll k$ (i.e. $p_\bot=|\textbf{p}_\bot|\ll 1$) and $\Psi_0(\textbf{x})=e^{ikn_0z} \psi(\textbf{x})$, where $\psi$ depends weakly on $z$. Thus, neglecting the paraxial derivative of $\psi$, we note that $p_z=n_0$, as seen immediately by the application of $(-ik^{-1}\partial_z)$ on $\Psi_0$. Equation (\ref{dirac}) then reduces to the Dirac-like eigenvalue equation
\begin{equation}
H\psi=-p_z\psi. \label{b}
\end{equation}
The first and the fourth components of $\psi$, to be denoted by $\psi_\pm$ respectively (see (\ref{0})), represent the two circularly polarized states. Equation (\ref{b}) is the Dirac-like paraxial wave equation for monochromatic paraxial propagation in weakly inhomogeneous media. It will be used in our study of the spin transport problem in the following sections.

\section{The Foldy-Wouthuysen transformation}

In homogeneous media ($\zeta=0$, $H\equiv H_0$), the above Hamiltonian resembles the free Dirac Hamiltonian. Like the Dirac Hamiltonian, $H_0$ has an odd (off-diagonal) part $O$ which involves $\textbf{M}_\bot$ and thus couples the circularly polarized states with the $z$-component of the electromagnetic field $F_z$ (see (\ref{3})). We can use the Foldy-Wouthuysen technique to construct a matrix $U$ that diagonalizes $H_0$ via the similarity transformation $U^{-1}H_0U$. In this scheme $U$ is given by $e^{\beta O \theta}$, or
$$
U(\textbf{p}_\bot)=\exp (\textbf{M}_\bot\cdot\textbf{p}_\bot\theta)=\cosh(p_\bot\theta)+\frac{\textbf{M}_\bot\cdot \textbf{p}_\bot}{p_\bot}\sinh(p_\bot\theta)
$$
having used (\ref{a}). ($U$ is Hermitian and not unitary as in the Dirac theory.) The similarity transformation diagonalizes $H_0$, provided
$$
\tanh(2p_\bot\theta)=\frac{p_\bot}{n_0}
$$
so that $U^{-1}H_0U=-E\beta$ and
\begin{equation}
U(\textbf{p}_\bot)=\frac{n_0+E+\textbf{M}_\bot\cdot\textbf{p}_\bot}{\sqrt{2E(n_0+E)}} \label{c}
\end{equation}
where $E=\sqrt{n_0^2-\textbf{p}_\bot^2}$. Under this transformation, the paraxial wave equation (\ref{b}) thus becomes 
\begin{equation}
H^\prime\psi^\prime=-p_z\psi^\prime \label{ab}
\end{equation}
where $\psi^\prime=U^{-1}\psi$ and
$$
H^\prime=U^{-1}HU=-E\beta-U^{-1}(\textbf{p}_{\bot})\beta\zeta(\textbf{x})U(\textbf{p}_{\bot}).
$$
The last term on the left hand side of the above can be written in the momentum representation as
\begin{eqnarray}
U^{-1}(\textbf{p}_\bot)\beta U(\textbf{p}_\bot)U^{-1}(\textbf{p}_\bot)\zeta(ik^{-1}\nabla_{\textbf{p}})U(\textbf{p}_\bot)=
\beta U^2 \zeta(ik^{-1}\nabla_{\textbf{p}}+k^{-1}\textbf{A}_\bot) \nonumber \\
=\frac{1}{E}[n_0 \beta+i\textbf{M}_\bot \cdot (\hat{\textbf{z}} \times \textbf{p}_\bot)]\zeta(\textbf{x}+k^{-1}\textbf{A}_\bot) \label{d}
\end{eqnarray}
where
$$
\textbf{A}_\bot({\textbf{p}_\bot})=iU^{-1}(\textbf{p}_\bot)\nabla_\textbf{p}U(\textbf{p}_\bot)=iU^{-1}(\textbf{p}_\bot)\nabla_{\textbf{p}_\bot}U(\textbf{p}_\bot).
$$
In deriving (\ref{d}), we have made use of the fact that $U^{-1}\beta=\beta U$ as well as the identity
$$
G^{-1}(x)f(\partial_x)G(x)=f(\partial_x+G^{-1}\partial_x G).
$$  
Direct calculation via (\ref{c}) yields
$$
\textbf{A}_\bot({\textbf{p}_\bot})=\frac{\beta\hat{\textbf{z}}\times\textbf{p}_\bot}{2E(n_0+E)}+\frac{i(\textbf{M}_\bot\cdot\textbf{p}_\bot)\textbf{p}_\bot}{2E^2(n_0+E)}+\frac{i\textbf{M}_\bot}{2E} .
$$
In paraxial propagation, we have $E\approx n_0=p_z$, as $p_\bot\ll (1 \le)n_0$. Collecting results, we therefore have,
\begin{eqnarray}
\psi^\prime=(1-\frac{1}{2p_z}\textbf{M}_\bot\cdot\textbf{p}_\bot)\psi \ \ \ \ \ \ \ \ \ \ \ \ \ \ \ \ \ \ \ \ \nonumber\\
H^\prime=-\beta n(\textbf{x}+k^{-1}\textbf{A}_\bot)-\frac{i}{p_z}\textbf{M}_\bot \cdot (\hat{\textbf{z}} \times \textbf{p}_\bot)\zeta(\textbf{x}+k^{-1}\textbf{A}_\bot) \label{q}
\end{eqnarray}
where
$$
\textbf{A}_\bot=\frac{\beta\hat{\textbf{z}}\times\textbf{p}_\bot}{4p_z^2}+\frac{i(\textbf{M}_\bot\cdot\textbf{p}_\bot)\textbf{p}_\bot}{4p_z^3}
+\frac{i\textbf{M}_\bot}{2p_z} .
$$

\section{Adiabatic evolution: The Berry effect terms}

As mentioned before, terms involving $\textbf{M}_\bot$ couple the circularly polarized states $\psi_\pm$ with the $z$-component of the electromagnetic field (which is, after all, small in view of paraxial propagation). Such terms are to be ignored in the adiabatic evolution, which means that we can project $H^\prime$ on the polarization subspace, keeping only the corresponding elements (elements 11,14,41,44). Equations (\ref{q}), therefore, read:
$$
H^\prime=-\sigma_z n(\textbf{x}+k^{-1}\textbf{A}_\bot),  \ \ \ \psi^\prime=\left( \begin{array}{c}\psi_+\\ \psi_- \end{array} \right) 
$$
where $\sigma_z$ is the Pauli matrix and
$$
\textbf{A}_\bot=\frac{\sigma_z\hat{\textbf{z}}\times\textbf{p}_\bot}{4p_z^2}=\frac{\sigma_z\hat{\textbf{z}}\times\textbf{p}}{4p_z^2}.
$$
 The paraxial wave equation (\ref{ab}), thus, becomes
$$
{\cal H} \psi_\sigma=\sigma p_z \psi_\sigma
$$
where $\sigma=\pm $ represents the polarization and 
\begin{equation}
{\cal H}(\textbf{x},\textbf{p})=n(\textbf{x}+k^{-1} \sigma{\bf {\cal A}}_\bot), \ \ \ {\bf {\cal A}}_\bot=\frac{\hat{\textbf{z}}\times\textbf{p}}{4p_z^2}. \label{eq}
\end{equation}
The Hamiltonian ${\cal H}$ governs the adiabatic evolution of polarized paraxial waves. It contains the `spin-orbit' interaction via $\sigma{\bf {\cal A}}_\bot$, as a result of which, the position operator acquires an anomalous contribution according to $\textbf{x}\rightarrow\textbf{r}=\textbf{x}+k^{-1} \sigma{\bf {\cal A}}_\bot$. The physical (observable) beam coordinates are no longer the canonical coordinates $\textbf{x}=ik^{-1}\nabla_\textbf{p}$; they are now $\textbf{r}=ik^{-1}{\cal D}_\textbf{p}$, where ${\cal D}_\textbf{p}=\nabla_\textbf{p}-i\sigma{\bf {\cal A}}_\bot$ is the covariant derivative. The spin-orbit interaction, thus, introduces a Berry connection (gauge potential) ${\bf {\cal A}}_\bot$ in the momentum space. This is in perfect analogy with the electromagnetic interaction, with the role of the position and momentum interchanged (and helicity replaced by charge, of course) and is a well-known feature of spin transport \cite{Bliokh,Bliokh1,Bliokh2,Berard,Mehrafarin}. The physical coordinates are non-commutative: 
$$
[r_i,r_j]=ik^{-2}\sigma \varepsilon_{ijk}B_k
$$
where 
$$
\textbf{B}=\nabla_\textbf{p}\times{\bf {\cal A}}_\bot=\frac{\textbf{p}}{2p_z^3}\;(=\frac{\textbf{p}}{2p^3})
$$
is the Berry curvature (gauge field strength) associated with the connection. (In the last expression, which corresponds to a magnetic-monopole-like field, we have used $p=\sqrt{p_z^2+p_\bot^2}\approx p_z$.)

\section{Spin Hall effect and the Rytov law}

In geometrical optics, the ray equations can be obtained \cite {Kratsov} via the Hamilton's equations 
$$
\dot{\textbf{p}}=-\nabla_\textbf{\scriptsize x}{\cal H}, \ \ \ \dot{\textbf{x}}=\nabla_\textbf{\scriptsize p}{\cal H}
$$
where dot denotes derivative with respect to a beam parameter, which, here, corresponds to the beam length. The canonical variables $\textbf{x}=\textbf{r}-k^{-1}\sigma {\bf {\cal A}}_\bot$ and $\textbf{p}=k^{-1}\textbf{k}$ are considered classical, of course. Along the beam, $\textbf{r}$ represents the position and \textbf{p} the (unit) wave vector. For paraxial transport, using (\ref{eq}), we thus have
\begin{equation}
\dot{\textbf{p}}=-\nabla_\textbf{\scriptsize r} n(\textbf{r}), \ \ \ \dot{\textbf{r}}=k^{-1}\sigma\textbf{B}\times \dot{\textbf{p}} \label{g}
\end{equation}
which, of course, reduce to the standard ray equations in the zeroth approximation $k^{-1} \rightarrow 0$. In this (`classical') limit, the spin-orbit interaction vanishes and the right/left circularly polarized rays follow the same trajectory, which is a straight line along the paraxial direction. In the first (`semiclassical') approximation, however, these rays split due to the effect of the Berry curvature according to (\ref{g}). The deflections from the paraxial trajectory are given by
$$
\delta\textbf{r}_\bot=k^{-1}\sigma\int_C \frac{\textbf{p}\times d\textbf{p}}{2p_z^3}
$$
where $C$ is the beam trajectory in momentum space. The resulting displacements, being in opposite directions, are orthogonal to the paraxial direction and produce a spin current across the direction of propagation. This is the spin Hall effect for paraxial beam transport.

The phase change suffered by the beam in the course of its propagation is given by
$$
\phi=-\omega t+\int \textbf{k} \cdot d\textbf{x}=-\omega t+ \int \textbf{k} \cdot d\textbf{r}+\sigma\int_C {\bf {\cal A}}_\bot\cdot d\textbf{p}
$$
where, in the last integral we have used the fact that ${\bf {\cal A}}_\bot\cdot \textbf{p}=0$. This integral represents the geometric Berry phase, which is of opposite signs for the two polarizations. For a linearly polarized beam, it therefore leads to the rotation of the polarization plane through the angle 
$$
\int_C {\bf {\cal A}}_\bot\cdot d\textbf{p}=\int_C \frac{1}{4}\tan^2\theta\; d\varphi \approx\frac{1}{4}\int_C \theta^2\; d\varphi
$$
where $\theta(\ll 1)$, $\varphi$  are the zenith and azimuthal angles of the beam, respectively. This is the Rytov law for paraxial beam transport.

\end{document}